\documentclass{article}
\usepackage[affil-it]{authblk}
\newtheorem{theorem}{Theorem}
\input epsf 
\title{A Note on Graphs with 2 Bends}
\author{Martin Pergel\thanks{Partially supported by a Czech research grant GA\v{C}R GA14-10799S.}}
\affil{Department of Software and Computer Science Education,\\
Charles University,\\Malostransk\'e n\'am. 25,\\ 118 00~~Praha 1,\\ Czech Republic.}

\begin{document}
\def\Proof{\par{\noindent\bf Proof. }}
\maketitle
\def\fixme#1{{\bf FIXME!!!} #1\par}
\begin{abstract}
We show NP-completeness for the recognition problem of classical
2-line-bend graphs.
\end{abstract}
Bend-number is an interesting topic for graphs with geometrical
representations. Recently classes of $k$-line-bend graphs were explored
\cite{Izraelci,Nemci} after being defined already in 1980s \cite{Izraelci2}
and plenty of properties got proven.

The aim of this preprint is to share the proof with colleagues who have asked me about this proof before it gets published with further results. Thus we expect the reader to know the problem and also we expect the reader to understand the motivation of this problem.

For these graph-classes, it is known
that 0-line-bend graphs can be recognized in polynomial time (as these are
interval graphs, i.e., intersection graphs of intervals on a line, see, e.g.,
\cite{Kratsch})
and that the recognition of 1-line-bend graphs is NP-complete \cite{Nemci}.
It is also known that given prescribed sets of 1-bend objects, the recognition
problem is still NP-hard \cite{Steve}.

\section{2-bend graphs}
In this section we show the following:
\begin{theorem}
It is NP-complete to decide whether a given graph has 2-line-bend representation.
\end{theorem}

\Proof
The NP-membership is obvious. As a polynomial certificate we use a list
of coordinates denoting start- and end-points of axis aligned straight-line
segments. Such a representation has polynomial size w.r.t. the given graph.

For the NP-hardness we use a polynomial reduction from PURE-NAE-3-SAT.
This problem is derived from classical 3-SAT (satisfiability of formula
in conjunctive normal form where each clause has at most 3 literals).
NAE-3-SAT asks for truth-assignment where neither all three literals are
true nor false. PURE-NAE-3-SAT is a version without negations (i.e., all
variable-occurences in the formula are positive). PURE-NAE-3-SAT is
a well-known NP-complete problem whose NP-hardness witnesses the problem
of bicoloring 3-uniform hypergraph.

\begin{figure}
\epsfbox{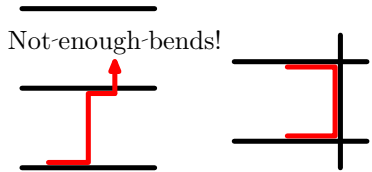}
\caption{It is impossible to intersect three parallel segments with a 2-bend
curve while two parallel and one perpendicular can be intersected.}
\label{aa:claumean}
\end{figure}

First we present the idea of the reduction, then we focus on technical details.
The idea is that each variable-occurence gets represented by a vertex and in the
line-bend representation it turns up as either vertical or horizontal segment.
Vertical segments correspond to the assignment TRUE, horizontal segments
correspond to the assignment FALSE (thus it is necessary to keep all the
occurences synchronized). Each clause gets also represented by
a vertex adjacent to the appropriate vertices. Considering a representation
of variables by straight-line segments such that no pair of these segments
lies on a common line, note that it is impossible to represent the appropriate
clause-gadget and, conversely, it is possible to represent, both, a clause
with one literal true and two false and a clause with one literal false
and two true, see Figure \ref{aa:claumean}. 

\begin{figure}
\epsfbox{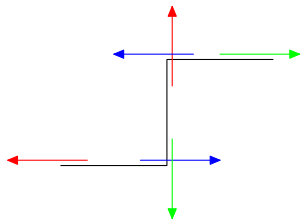}
\caption{The picture shows how 6 segments may exit a particular 2-bend curve
withouth having to bend inside it. All the other 2-bend curves non-intersecting
with these 6 ones must bend to get outside the black 2-bend curve.}
\label{aa:6exits}
\end{figure}

The construction uses two types of tricks. One type of tricks is
a "quantitative trick" and it uses the fact that when a particular problem
may occur only constantly many times in the whole construction (when
representing the formula), we do not have to care about it and we bypass
the problem by repeating the whole formula several times (we could also
add sufficiently large non-satisfiable part of the formula and force the
construction to use those "singularities" to represent the artificial
non-satisfiable part of the formula). A typical representant of the problems
solved by this trick is that considering a representant of a vertex adjacent
to many other mutually non-adjacent vertices (e.g., a vertex adjacent to all
vertces representing variables), only constantly (namely at most 6) many
among the other vertices may leave the representation of the main vertex
without using an extra bend for it (they pass through the endpoints of
individual segments), see Figure \ref{aa:6exits}.
Also some variable-representants
may be cheating if the representants of $a$ and $b$ mutually intersect, but
it is also a constant number.
The other type of tricks are technical tricks (that are either necessary
for enforcing the properties of the construction or that are enforced by
geometrical structure of the recognition problem, e.g., intersections
or non-intersections of clause-representants).

The "quantitative trick" gets sometimes used implicitly and usually we won't
be determining precisely how many times it is necessary to replicate the
formula to waste these unwanted properties.

After explaining the idea, we start describing the technical background
of the construction. First point is, how to represent individual occurences
just by vertical or horizontal segment and how to keep the occurences
synchronized: This gets done by representing variables and occurences.
The representant of a variable is responsible for the synchronization
of the occurences. Each variable, in general, get represented (in the
graph) an individual vertex attached to special vertices $a$ and $b$.
Vertices $a$ and $b$ get represented by at most twice bending sequence
of segments and in order to intersect this pair of curves by the curves
representing the variables, at most 12 representants of the variables may
avoid wasting one bend for intersecting $a$ and also one for intersecting
$b$. Therefore the first and the third segment in their representation will
be covered by one of the segments representing $a$ and $b$. In this way
we obtain the representation of variables. Those 12 exceptional variables
get solved by replicating the formula 13 times ("quantitative trick").

For a variable $v$ we represent its occurences $v_1, v_2, ...$ by vertices
adjacent to $a, b$ and $v$. Note that in order to represent intersection
with $a$ and $b$ we need two segments (that get covered by segments
representing $a$ and $b$) and yet we have to intersect $v$. How to do this?
We have to intersect $v$ simultaneously with either $a$ or $b$. This
(together with the fact that first and third segment are wasted by intersection
with $a$ and $b$, respectivelly) forces the occurences of a particular
variable to have the vertical-horizontal orientation of the middle-segment
to be synchronized with the orientation of the middle-segment of the whole
variable.

The clauses get represented by vertices adjacent to the appropriate occurences
of the appropriate (three) variables. The representants of the clauses in
an arrangement where no pair of occurence-vertices is represented by curves
having the middle-segment on a common line, it is impossible to represent
an unsatisfied clause.

\begin{figure}
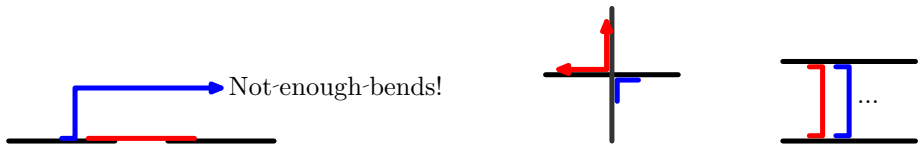

\epsfbox{bendy.3}\hfill\epsfbox{bendy.4}\hfill\epsfbox{bendy.5}
\caption{{\bf Left:} We cannot intersect two colinear segments by two
mutually non-intersecting 2-bend curves -- the blue 2-bend curve has no further
bends to reach the right segment.
{\bf Middle:} How to intersect two mutually intersecting segments by a pair
of mutually non-intersecting 2-bend curves. {\bf Right:} Non-colinear
parallel pair of segments can be intersected by arbitrarily many mutually
non-intersecting 2-bend curves.}
\label{aa:dvespojky}
\end{figure}

Now, it remains to add the technical details that disallow two occurences
from the same clause to have the middle-segment on a common line. Note that
such a situation for a pair occurence-vertices $v_i$ and $u_j$ (whose
middle-segments should occur on a common line) is incompatible with adding
a pair of non-adjacent vertices when both newly added vertices shall be
adjacent to both occurence-vertices, i.e., $v_i$ and $u_j$. Also note
that if $v_i$'s middle segment was vertical and $u_j$'s horizontal, we
couldn't represent the clause-vertex now. Thus instead of a pair of vertices
we add just one of them and instead of the second one we use the representation
of the clause, see Figure \ref{aa:dvespojky}. 

\begin{figure}
\epsfbox{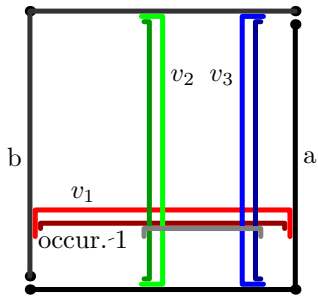}
\caption{How to obtain a representation from a satisfiable formula. Picture
shows three variables $v_1, v_2$ and $v_3$ each with one occurence (darker
color) and a representation of a clause (grey).}
\label{aa:global}
\end{figure}

Now, it is clear that for an unsatisfiable formula we cannot obtain
a representation (as the variable-representants cannot be divided into
vertical and horizontal ones in such a way that each clause contains
at least one vertical and at least one horizontal). What remains is to
show that the satisfiable formula can always be represented. For the
explanation we use Figure \ref{aa:global}. 
Variables evaluated to "true" we represent
as the vertical ones, variables evaluated to "false" we represent horizontally.
Individual occurences get represented in the close neighborhood of the
appropriate variables. Representants of the clauses should now intersect
such a triple of segments that w.l.o.g., two are horizontal and one vertical.
Moreover, the vertical one intersects both horizontal ones. Thus we obtain
the appropriate representation by picking the (sub)segment of the vertical
segment between the intersections with the horizontal ones and to such
a segment we add small particles passing along both horizontal segments.
Yet it remains, how to represent the "colinearity-obstructions". For the
mutually intersecting pairs of segments (i.e., vertical against horizontal)
we represent it opposite to the variable-representation, see Figure \ref{aa:dvespojky}, the middle picture.
For a pair of w.l.o.g. vertical segments, we attach them
by a horizontal segment and we add two small vertical segments inside those
two vertical segments. Note that it is simple to avoid representation of
clauses as on each segment at most one representation of a clause appears
and none of those clause-representants appears near "the end of that segment",
i.e., near the representation of extra vertex $a$ and $b$ (mentioned at the
beginning of the construction).
%\end{proof}

\end{document}